\documentstyle[aps,twocolumn]{revtex}

\begin{document}
\title{An ultrahigh-$Q$ microsphere laser based on the evanescent-wave-coupled gain}
\author{Yong-Seok Choi, Hee-Jong Moon$\dagger$, Sang Wook Kim, and Kyungwon An}
\address{Center for Macroscopic Quantum-Field Lasers and Department of Physics,\\
Korea Advanced Institute of Science $\&$ Technology, Taejon 305-701, Korea}

\date{\today}
\maketitle
                             
\begin{abstract}                              
{We have demonstrated an ultrahigh-$Q$ whispering-gallery-mode (WGM) microsphere laser based on the evanescent-wave-coupled gain. Dye molecules outside the sphere near the equator were excited, resulting in  WGM lasing in the lowest radial mode order. The loaded quality factor of the lasing WGM was $8(2)\times 10^9$, the highest ever achieved in the microlaser.}
\end{abstract} 

\vspace{-0.35in}
The potential of fused silica microspheres lies in the existence of the resonator modes of ultrahigh cavity quality factor $Q$ in the optical region.  These modes, known as whispering gallery modes (WGM's), originate from the total internal reflections of optical waves inside the sphere. Recent $Q$ measurements of microspheres show that under proper conditions the bare cavity quality factor can be as high as $8\times10^9$ \cite{Gorodetsky-OL96,Vernooy-OL98}.
In addition to ultrahigh $Q$, low-order WGM's have much smaller mode volume than the other types of cavities of the same dimension, thus enabling much stronger matter-field coupling strength \cite{cavityQED}.  Based on these features, host of applications have been envisioned and some have been demonstrated in the various fields of optical science such as laser frequency stabilization \cite{Schiller-OL91}, highly efficient optical power coupler \cite{Cai-PRL00}, thresholdless microlaser \cite{Yamamoto-PT93} and the cavity quantum electrodynamics studies of atoms \cite{Vernooy-PRA98}, molecules \cite{Norris-APL97}, semiconductor nanocrystals \cite{Brun-PRA00} or quantum dots \cite{Pelton-PRA98,Fan-OL99}. However, the loaded $Q$ or the actual $Q$ of microsphere in these applications has been so far much lower than those of the bare microspheres, only in the range of $10^{6-7}$due to many technical reasons \cite{Brun-PRA00,Gorodetsky-JOSAB99,Treussart-EPJD98}, thus making most of the aforementioned applications still elusive. 

Recently, we have reported the success of the microcylinder laser based on the evanescent-wave-coupled gain \cite{Moon-PRL00}. In this system, the gain for laser oscillation comes only from the evanescent-wave coupling between the gain molecules outside and the high-$Q$ WGM's inside.  This type of coupling configuration is preferable in the microcavity lasers with ultrahigh $Q$.  
In the usual WGM microcavity laser, where the cavity itself is the gain medium, the cavity $Q$ tends to be degraded due to the thermal stress imposed by direct 
electric or optical excitation in the cavity.  However, in our configuration, such degradation can be avoided since the medium outside the cavity is excited, leaving the cavity unaffected by the pumping. In our microcylinder laser, the loaded $Q$ reached as high as $3\times 10^7$.  
Considering that our cavity was nothing but a segment of ordinary multi-mode fiber with its jacket removed, we believe that the observed loaded $Q$ must be very close to the ultimate bare $Q$ of this type of cavity.  
This observation motivated us to apply the same configuration to a microsphere whose cold $Q$ can be in the range of $10^{9-10}$ and to investigate how high actual $Q$ can be achieved in WGM lasing.

In this Letter, we report realization of an ultrahigh $Q$ microsphere laser of the lowest WGM mode order whose loaded cavity quality factor is $8(2)\times 10^9$ {\em when the WGM lasing actually occurs in the cavity.}  
The overall quality factor, $Q_t$, representing the {\rm total} loss in the lasing WGM, is found to be $4(1)\times$ 10$^{9}$. 
The loaded $Q$ value reported in this work is the highest ever observed so far in all the applications using microcavities.  In addition, our results demonstrate the feasibility of a single quantum-radiator microsphere laser. 

In our experiments, a microsphere of 102(1) $\mu$m diameter was surrounded by a gain medium, ethanol (refractive index $n_1\simeq$\,1.361) doped with Rhodamine 6G (Rh6G) dye molecules as shown in Fig.\,1. 
The microcavity used in our experiment was a microsphere fabricated by melting the tip of a fused silica fiber with two counter-propagating CO$_2$ laser beams \cite{Weiss-OL95}. The typical diameter at the tip was 15 $\mu$m. 
By controlling the heating time and/or the laser power, we could control the size of microspheres in the accuracy of a few microns. The refractive index $n_2$ of the sphere was 1.471.
Since the $Q$ value of the present microsphere was expected to be in the range of $10^{9-10}$, much higher than that of the microcylinder of Ref.\cite{Moon-PRL00}, the concentration of Rh6G was made dilute, even down to 0.0125mM/L, which is only 1/160 of the density in our previous microcylinder laser.  In this way we can ensure that only the WGM with the lowest loss (or the highest $Q$) can undergo laser oscillations. The gain medium was contained in a fused silica capillary of 200 $\mu$m inner diameter and 320\,$\mu$m outer diameter. The microsphere was inserted in the dye filled capillary right after its fabrication in order to prevent any surface contamination by water moisture and dust particles. 
As a pumping source, we used a 532\,nm-pulsed excitation of a frequency-doubled $Q$-switching Nd:YAG laser (Quantel Brilliant $\omega$) with pulse width of 10 ns and repetition rate of 10 Hz. The pump laser of about 1 mm beam waist with its polarization parallel to the capillary axis (let us define it $z$ axis) was focused near the equator of the sphere with a cylindrical lens in order to excite the gain molecules there, and thus to excite only the WGM's near the equatorial plane. The beam full width along the $z$ axis at the sphere was about 20 $\mu$m. Since the dye concentration was very dilute, the pump beam intensity did not change much throughout the medium.
We measured the emission spectrum with a spectrometer consisting of a 1/4-meter spectrograph (MS257) and a computer controlled photodiode array (PDA) detection system (Instaspec II). The PDA was triggered by the pump laser with exposure time of 32 ms so as to ensure the complete recording of the spectrum resulting from a single pulse excitations. Color filters were used to block the intense pump light entering  the spectrometer and the microscope as shown in Fig.\ref{fig1}.

A typical image of laser oscillation observed with the microscope is shown in  Fig.\,\ref{fig2}, where the WGM lasing light emerges tangentially from the surface of microsphere near the equator (in red). In this case, the pumping energy was about 200 $\mu$J and the dye concentration was 0.05 mM/L.  The lasing spectrum is shown in Fig.\,\ref{fig3}(a), where a prominent WGM peak appears with other smaller ones near at 611 nm. These WGM's are longitudinal modes of the same mode order as discussed below. The observed mode spacing, 0.79 nm, is consistent with the calculated mode spacing. 
Polarization analysis reveals that these peaks have the polarization parallel to the $z$ axis (i.e., TE polarization). 

WGM in a spherical cavity is specified by the radial ($r$) mode order $\nu$, the azimuthal (in the $\phi$ direction) mode number $l$, the polar ($\theta$) mode number $m$ and the polarization (TM or TE) \cite{Chang-90,Barber-90}. The WGM's excited in the experiment are of the largest $|m|\simeq l$ since they are confined near the equator.  Only the molecules residing in the evanescent-field region of the WGM can participate in the laser oscillation.  We define the occupation factor $\eta$ as the fraction of the evanescent field volume to that of the whole WGM.  Then, as in Ref.\,\cite{Moon-PRL00}, assuming a four-level system for the dye molecules, the laser threshold condition can be written as
\begin{equation}
N_1 \eta \sigma_e(\lambda)~\ge~N_0 \eta \sigma_a(\lambda)~+~{2 \pi n}/{\lambda Q}\;
\end{equation}
where $N_{0}$ and $N_{1}$ are the number density of the dye molecules in the ground electronic singlet state and in the lowest excited singlet state, respectively, $\sigma_e(\lambda)$ and $\sigma_a(\lambda)$ the emission and absorption cross section of the dye molecules at wavelength $\lambda$, respectively, $n$ is the relative index of refraction defined by $n_2/n_1$, and $Q$ is the loaded cavity quality factor excluding the dye absorption loss. This threshold condition leads to the minimum fraction of excited molecules, $\gamma(\lambda) \equiv N_1 / N_t$, with $N_{t}$ the total number density of the molecules, for inducing laser oscillation at $\lambda$. 
The lower $\gamma(\lambda)$ is, the weaker the threshold pump power becomes at the given lasing wavelength $\lambda$. 
As the pump power increases, WGM lasing peaks of a given mode order appear around the minimum wavelength of $\gamma(\lambda)$ curve corresponding to that WGM mode order.

As discussed in Ref.\,\cite{Moon-PRL00}, the minimum of $\gamma(\lambda)$ shifts to the red as $N_t (\simeq N_0 )$ or $Q$ increases. This trend is demonstrated in Fig.\,\ref{fig3}(b), where the same microsphere was repeatedly used in three different concentration settings. 
For different dye concentrations, the laser oscillations were found at different wavelengths in the spectrum \cite{Mazumder-OL95}.  Lasing was observed even with a very low concentration of 0.0125 mM/L around 598 nm with the mode spacing of 0.75 nm. For the concentration of 0.2 mM/L, the number of lasing WGM's was larger than in the other two cases and the WGM peaks appeared in a broad range around 615 nm, which is consistent with the broad minimum of $\gamma(\lambda)$ curve in this case.

Comparing the location of the minimum of $\gamma(\lambda)$ for various values of fitting parameter $Q'\equiv \eta Q$ with
the center wavelength of the each WGM group in Fig.\ \ref{fig3}(b), we can find the loaded $Q$ associated with each WGM group. Fig.\ \ref{fig4} shows the calculated $\gamma(\lambda)$ curves which best fit the center wavelengths of the observed WGM groups. 
For this fitting we actually measured the absorption and emission cross sections of Rh6G molecules. Particularly, since the dye absorption is very weak in the wavelength region $\ge$ 600 nm, we measured the absorption through a 15 cm long dye cell filled with high-concentration(4 mM/L) Rh6G solution in ethanol. 
In the case of 0.05mM/L (0.2 mM/L), the mode group centered around 611 nm (615 nm) can be fitted with $Q$ = 8(2) $\times$ 10$^9$ ($Q$ = 5(2) $\times$ 10$^9$) with $\eta\simeq 1/36$, which is numerically calculated for the $\nu=1$ modes (see the discussion below) from the elastic scattering theory \cite{Barber-90}. 
The peaks centered at 598 nm for $N_t$ = 0.0125\,mM/L can be fitted with $Q$ = 1.9(7)$\times$ 10$^9$ with $\eta \simeq 1/37$. 
The $\eta$ values used in our fittings were calculated for absorption-free microspheres. From numerical simulations with a size parameter smaller than the actual one we found that the value of $\eta$ slightly decreases when the absorption loss is included. This trend would make our fitted $Q$ values slight larger, but still within the experimental error. In our fittings listed above, we neglected this effect.

The large variation in fitted $Q$ is attributed to gradual degradation of actual $Q$ in the course of three successive experiments and possibility of exciting different WGM's in each experiment.  We used the same microsphere in the three experiments in the order of 0.05 mM/L, 0.0125 mM/L and 0.2 mM/L. Before trying a new concentration, the capillary and the microsphere were flushed with ethanol. Each experiment took about an hour. Interestingly, very high $Q$ values could be maintained over an extended period of time in the ethanol environment, contrary to the fact that even the bare $Q$ values in previous studies lasted in air only for a fraction of the time span reported in this work. \cite{Gorodetsky-OL96,Vernooy-OL98}. 

One can define the absorption quality factor, $Q'_{\rm abs}\equiv 2\pi n/(N_t \eta \sigma_a \lambda)$, corresponding to the dye absorption loss in the evanescent field region.  Since $N_0 \simeq N_t$, the right hand side of Eq.\ (1) is summarized by an overall quality factor, $1/Q_t \equiv 1/Q + 1/Q'_{\rm abs}$, representing the {\rm total} loss in the lasing WGM's.  In the above experiment, $Q_t \simeq 4(1)\times 10^9$ for the 611 nm lasing. 

From the cavity ringdown of the WGM lasing signal we also directly measured the loaded $Q$. 
For this experiment a beam splitter was introduced inside the spectrometer so that the spectrum and the time decay of the lasing signal could be measured simultaneously.  
The experiment was done with individual microspheres with their $Q$ values ranging from $10^7$ to a few times of $10^9$. 
The decay curves were recorded with a photomultiplier tube (Hamamatsu R647-p) connected to a fast digital oscilloscope (Lecroy 9370C) and averaged over 500 shots.  The decay signal consists of two exponentials, i.e., a fast fluorescence decay (decay time$\sim$10 ns) with large amplitude and a slow WGM decay with a small amplitude. We used a 50 $\mu$m slit at the imaging plane of the PMT port in order to limit the spectral contents going into the PMT along with the lasing signal of a narrow linewidth. Under this condition the amount of the fluorescence photons were comparable to that of the lasing photons, but concentrated within about 10 ns time slot in the beginning of the curve. Particularly, when $Q$ was larger than $2\times 10^9$ the slow decay part was difficult to detect since the same amount of photons per shot spread over a longer decay time.  For low $Q$ values ($\sim 10^7$) the WGM decay component became as fast as the fluorescence decay, so our two exponential fit could give a large error. 
Nonetheless, we could still isolate the WGM decay component. 
The loaded $Q$ values obtained from the time decay data are fairly consistent with those from the spectrum data as seen in Fig.\ \ref{fig5}.  This consistency strongly supports our assessment of $Q$ values in Fig.\ \ref{fig4}.

The loaded $Q$ values measured in our experiments are as high as the expected bare $Q$ values at the wavelengths around 600 nm.
According to the elastic scattering theory \cite{Barber-90}, the ideal quality factors for the TE polarized WGM's around 598 nm are $1.8 \times 10^{11}$ and $6.2 \times 10^7$ for $\nu$ = 1 and 2, respectively. It is obvious that the observed $Q$ values cannot be assigned to $\nu$ = 2 modes.  These $Q$ values should correspond to $\nu$ = 1 modes with $Q$ degradation in actual experiments.  The major $Q$-degradation in our spheres comes from the absorption loss in the sphere medium (fused silica) \cite{Vernooy-OL98}. The $Q_{\rm med}$ due to the absorption loss in the microsphere medium is $7\times10^9$ at around 600 nm \cite{Vernooy-OL98}. The other sources of $Q$ degradation are the surface scattering loss of microsphere and the absorption loss in the ethanol solution, but these effects are negligible as shown below.
We measured the surface roughness of our spheres using the atomic force microscopy and obtained about 0.1(1) nm rms with 3(2) nm correlation length, which correspond to $Q_{\rm sc} > 10^{12}$ at the lasing wavelength, based on the formula in Ref.\ \cite{Vernooy-OL98} with $\epsilon=n^2\simeq 1.17$.  In addition, the ethanol solution used in our experiments contains less than 0.2\% of water molecules in molar weight. These water molecules are uniformly distributed in ethanol. The absorption loss due to the water molecules in the evanescent-wave region of WGM at the wavelength of 600 nm corresponds to $Q_w \simeq 2\times 10^{11}$.
Therefore, we estimate the cold $Q$ of our microlaser to be $7\times10^9$, which is close to the loaded $Q$ values determined in our experiments.

Our results imply the feasibility of a single quantum-radiator microsphere laser. The volume of the evanescent-field region of the WGM with $\nu=1$ mode order is about $5\times 10^{-10}$ cm$^3$\cite{Vernooy-PRA98}, so that  the number of Rh6G molecules participating in the lasing was about $4\times10^6$ for the dye concentration of 0.0125mM/L. Note that due to the coherence dephasing of induced dipole the emission cross section of the dye molecules is reduced by a factor of 10$^{6-7}$ from that of a perfect quantum radiator such as a two-level atom or a quantum dot at low temperature \cite{Bawendi}. Therefore, the number of dye molecules in our experiment is roughly equivalent to a single ideal radiator. However, the feat of coupling such a single radiator to the evanescent region of WGM is yet to be demonstrated.

In conclusion, we have demonstrated a microsphere laser based on the evanescent-wave-coupled gain and found that the loaded $Q$ of the lasing WGM was as high as $8(2)\times 10^9$, only limited by the intrinsic absorption in the cavity medium.  Our results demonstrate a promising configuration of the ultrahigh-$Q$ microsphere in various scientific studies maintaining its high quality with negligible $Q$ degradation. 

This work was supported by the Creative Research Initiatives of the Korean Ministry of Science and Technology.\\

$\dagger$ Present address: Department of Optical Engineering, Sejong University, Seoul 143-747, Korea.

\begin{figure} 
%\vspace{3in}
%\special{epsf=fig1-small.eps scale=0.275} 
\caption{Experimental setup. $2a= 102(1) \mu$m, $2b=200 \mu$m, $n_1 \simeq 1.361$, $n_2 \simeq 1.471$. }
\label{fig1}
\end{figure}

\begin{figure} 
%\vspace{1.6in}
%\special{epsf=fig2-small.eps scale=0.75} 
%\vspace{0.15in}
\caption{The image of the microsphere (left) and its WGM lasing based on the evanescent-wave-coupled gain (right). The concentration $N_t$ was 0.05\,mM/L for the lasing image.}
\label{fig2}
\end{figure}

%\newpage
%\hbox{}

\begin{figure}
%\vspace{2.1in}
%\special{pict=fig3.pict scale=0.25} 
%\vspace{0.1in}
\caption{The emission spectrum of the ultrahigh-$Q$ microsphere laser (a) The pumping energy vs. the emission intensity with $N_t$ of 0.05mM/L. (b) The spectral shifts due to the change in the dye concentration.}
\label{fig3}
\end{figure}

\begin{figure}
%\vspace{2.5in}
%\special{pict=fig4.pict scale=0.31} 
%\vspace{0.1in}
\caption{Threshold population $\gamma(\lambda)$ curves which best fit the center wavelengths of the observed WGM groups in Fig.\ \protect{\ref{fig3}}(b).}
\label{fig4}
\end{figure}

\begin{figure}
%\vspace{1.4in}
%\special{pict=fig5.pict scale=0.19} 
%\vspace{0.1in}
\caption{(a) Correlation between the loaded $Q$ values found from the spectra and those from the ringdown of the WGM lasing light when they were simultaneously measured. (b) Ringdown signal of WGM lasing light corresponding to $Q \sim 1.3\times10^9$ (a decay time of 220(60) ns). The fast decaying part is due to the dye fluorescence. }
\label{fig5}
\end{figure}
 
\end{document}